\documentclass{iopjournal}

\makeatletter
\renewcommand{\articletype}[1]{%
  \noindent
  {\scriptsize\sffamily\bfseries\MakeUppercase{#1}}%
}
\makeatother

\makeatletter
\providecommand{\articletype}[1]{}
\makeatother

\usepackage{amsmath} 

\usepackage{amsmath}
\usepackage{booktabs}
\usepackage{tabularx}
\usepackage{array}

\begin{document}

\articletype{Article type} 

\title{Full-Potential Korringa–Kohn–Rostoker Study of Structural and Mechanical Properties in Si$_{1-(x+y)}$Ge$_y$Sn$_x$ Alloys \footnote{Notice: This manuscript has been coauthored by UT-Battelle, LLC, under Contract No. DE-AC0500OR22725 with the U.S. Department of Energy. The United States Government retains and the publisher, by accepting the article for publication, acknowledges that the United States Government retains a non-exclusive, paid-up, irrevocable, world-wide license to publish or reproduce the published form of this manuscript, or allow others to do so, for the United States Government purposes. The Department of Energy will provide public access to these results of federally sponsored research in accordance with the DOE Public Access Plan (\href{http://energy.gov/downloads/doe-public-access-plan}{http://energy.gov/downloads/doe-public-access-plan}).}}

\author{Wasim Raja Mondal$^1$\orcid{0000-0000-0000-0000}, Swarnava Ghosh$^2$\orcid{0000-0003-3800-5264}, Ka-Ming Tam$^{3}$, Markus Eisenbach$^{2}$\orcid{0000-0001-8805-8327}, Yang Wang$^{4}$\orcid{0000-0000-0000-0000}, and Hanna Terletska$^{1,*}$} 

\affil{$^1$Department of Physics and Astronomy, Middle Tennessee State
University, Murfreesboro, TN 37132 USA}

\affil{$^2$Advanced Computing for Chemistry and Materials, Oak Ridge
National Laboratory, Oak Ridge, TN 37831, USA.}

\affil{$^3$ 
Center for Computation and Technology,
Louisiana State University,
Baton Rouge, Louisiana 70803, USA"
}%

\affil{$^4$ Pittsburgh Supercomputing Center, Carnegie Mellon University,
Pittsburgh, PA 15213, USA.}

\affil{$^*$Author to whom any correspondence should be addressed.}

\email{Hanna.Terletska@mtsu.edu}


\begin{abstract}
Accurate first-principles descriptions of open, covalently bonded crystal structures remain challenging within conventional Korringa--Kohn--Rostoker (KKR) approaches owing to the spherical approximations commonly employed for the effective crystal potential. Here, we systematically assess the recently developed full-potential KKR (KKR-FP) implementation in the open-source MuST package and demonstrate its applicability to diamond-structured group-IV semiconductors and Si$_{1-(x+y)}$Ge$_y$Sn$_x$ alloys. Benchmark calculations for diamond-structured Si reveal substantial limitations of the conventional muffin-tin and atomic sphere approximations, whereas KKR-FP accurately reproduces the experimental equilibrium lattice parameter and bulk modulus, yielding $a_0=5.433$~\AA\ and $B_0=94.9$~GPa. The full-potential treatment also recovers the semiconducting character of Si. Further calculations for elemental Ge and $\alpha$-Sn establish the applicability of the approach across group-IV materials. Extending this framework to SiGeSn alloys, we investigate their composition-dependent equilibrium lattice parameters, bulk moduli, and mixing enthalpies. The calculated lattice parameters exhibit positive deviations from Vegard's law, while the bulk modulus decreases substantially with changes in alloy composition. Moreover, positive mixing enthalpies of $180.92$ and $346.32$~meV/atom are obtained for Si$_{0.5}$Ge$_{0.375}$Sn$_{0.125}$ and Si$_{0.375}$Ge$_{0.25}$Sn$_{0.375}$, respectively, indicating an energetic penalty for homogeneous alloy formation relative to the selected elemental reference states. These findings highlight the importance of nonspherical potential contributions in accurately describing open-structured semiconductors and establish KKR-FP within MuST as a promising framework for investigating the structural properties and mixing energetics of compositionally complex group-IV alloys
\end{abstract}

\section{Introduction}
An exact solution of the Schrödinger equation for solids remains a fundamental challenge in condensed matter physics, due to the complexity of its many-electron nature. Korringa \cite{KORRINGA1947392} and Kohn and Rostoker \cite{PhysRev.94.1111} provide a promising route for this problem by adopting multiple-scattering theory (MST) that laid the foundation for the Korringa--Kohn--Rostoker (KKR) method. Within this computational framework, the electronic structure of a solid is determined by describing electron scattering from individual atomic potentials and accounting for multiple-scattering processes throughout the crystal. One of the attractive aspects of the KKR formalism is that it yields a formally exact solution to the corresponding effective single-particle Schrödinger equation, subject to numerical convergence and approximations in representing the crystal potential.

Despite its rigorous theoretical foundation, the numerical implementation of the KKR method becomes computationally demanding, particularly when taking into account the full nonspherical character of the effective potential. To overcome this difficulty, several approximations have been developed to simplify the representation of the crystal potential while retaining the essential features of the electronic structure. Among these, the most widely employed approaches are the muffin-tin approximation (MTA) \cite{KORRINGA1947392,PhysRev.94.1111} and the atomic sphere approximation (ASA) \cite{PhysRevB.12.3060}. Other developments include the muffin-tin overlapping approximation (OMTA) \cite{Zwierzycki2009OMTA} and approaches that incorporate a spherically symmetrized charge density within the atomic sphere approximation \cite{PhysRevB.49.16694}. Although these approximations substantially reduce computational complexity, their accuracy is highly dependent on the crystal structure and the spatial variation of the effective potential.

In the MTA, the effective potential is assumed to be spherically symmetric within non-overlapping atomic spheres centered at the nuclei, while the potential in the interstitial region is approximated as a constant. This approximation is generally well suited to closely packed crystal structures, where the interstitial region occupies a relatively small fraction of the crystal volume. Despite its simplified description of the potential, MTA-based calculations have successfully reproduced several experimentally observed electronic properties, including band ordering, Fermi-surface features, pressure-dependent electronic behavior \cite{PhysRevB.8.1292}, and photoemission spectra \cite{PhysRevLett.41.339}, at relatively modest computational cost. Nevertheless, the approximation can introduce significant errors in calculated total energies and interatomic forces \cite{PhysRevB.46.12181}. Such limitations are particularly relevant for open, covalently bonded crystal structures, where directional bonding and nonspherical contributions to the effective potential play an important role \cite{Faulkner2005Scattering}. Consequently, conventional MTA-based calculations may become less reliable for these systems \cite{JIhm_1979,10.1063/1.1682462}.

The ASA provides an alternative simplification in which atomic spheres are expanded to approximately fill the crystal volume, allowing neighboring spheres to overlap and thereby avoiding an explicit treatment of the interstitial region. Although the ASA is computationally efficient and generally performs well for close-packed structures, its accuracy can deteriorate in open crystal structures with substantial interstitial regions. Several improvements have therefore been proposed to extend its applicability, including the introduction of empty spheres to provide a more complete representation of the crystal volume. For example, Glötzel \textit{et al}. \cite{GLOTZEL1980403} demonstrated improvements in ASA-based calculations for open-structure materials through an appropriate treatment of interstitial regions. Despite these developments, the spherical approximation to the effective potential remains a fundamental limitation, particularly when accurate structural properties, total energies, and interatomic forces are required. These considerations have motivated the development of full-potential KKR methods, which retain the nonspherical contributions to the effective potential and provide a more comprehensive description of the electronic structure \cite{DRITTLER199131,PhysRevB.55.4157,PAPANIKOLAOU1997131,PhysRevB.39.8187,PhysRevB.45.11527,PhysRevB.51.9560,PhysRevB.106.235114}.

Being a Green's function approach, the KKR method becomes a promising method for both ordered and disordered materials \cite{Ebertpsik}. The KKR-CPA approach enables the systematic investigation of chemically disordered systems like binary alloys \cite{PhysRevLett.26.253,PhysRevLett.41.339}, ternary alloys \cite{PhysRevB.93.054206}, and compositionally complex high-entropy alloys \cite{Troparevsky2015,OGURA2017454,KazunoriSato2023MT-MG2022012}. Beyond conventional electronic-structure calculations, KKR-based approaches have also been widely employed to investigate magnetocrystalline anisotropy \cite{OKUMURA2023115257}, chemical short-range order \cite{PhysRevB.102.054207}, and electronic transport properties \cite{turek2013electronic,PhysRevB.69.125104,PhysRevB.109.104204,10.1063/5.0065239}. Together, these developments have established the KKR Green's-function formalism as a powerful framework for investigating the electronic, structural, and magnetic properties of complex materials.

Recently, we developed a full-potential extension of the KKR Green's-function method, implemented in the open-source software package MuST \cite{Must}. In the present work, we employ this framework to investigate the structural and mechanical properties of Si$_{1-(x+y)}$Ge$_y$Sn$_x$ ternary alloys. These group-IV semiconductor alloys have attracted considerable interest because their structural and electronic properties can be tailored through compositional engineering, making them promising candidates for applications in infrared photodetectors \cite{MAEDA2024108304}, optical modulators, lasers \cite{Wirths2015,Seidel2024}, and photovoltaic devices \cite{10.1021/acsphotonics.6c00555}. 

Previous theoretical investigations of SiGeSn alloys mostly focus on electronic properties\cite{Ouserigha_2021,10.1063/1.5135324}. Recently, experimental investigations by Attiaoui \textit{et al}.\cite{attiaoui2026enhanced} have examined the stability of SiGeSn alloys, highlighting the importance of understanding the structural and thermodynamic consequences of alloying. We believe our calculations will help us to understand recent experimental observations\cite{attiaoui2026enhanced,10.1063/5.0151479,10.1021/acsami.2c14785,WIRTHS20161}.

\section{\label{sec:level1} METHOD AND CALCULATION DETAILS}
The Kohn-Sham (KS) equation\cite{PhysRev.136.B864, PhysRev.140.A1133} is the central equation in Density Functional Theory (DFT), which simplifies the many-body Schrödinger equation by mapping it to a system of non-interacting electrons moving in an effective potential. It is given by:
\begin{equation}
\Big\lbrack -\nabla^2 + V_{\text{eff}}([\rho(r)])\Big\rbrack \psi_i(r) = \epsilon_i \psi_i(r),
\label{eq:KS}
\end{equation}
where $\psi_i(r)$ is the wave function of the i$^{\text{th}}$ electron having eigenvalue $\epsilon_i$ under the effective potential $V_{\text{eff}}(r)$ that consists of the external potential $V_{\text{ext}}(r)$, the Hartree potential $V_{\text{H}}(r)$ and the exchange-correlation potential $V_{\text{XC}}(r)$. The exchange-correlation potential accounts for quantum effects that are not covered by the Hartree term. The Kohn-Sham equation must be solved self-consistently because 
$V_{\text{eff}}(r)$ depends on the electron density $\rho(r)$ which in turn depends on the $\psi_i(r)$ as given below
\begin{equation}
\rho(r) = \sum_{\epsilon_i\leq \epsilon_F}  |\psi_i(r)|^{2}  
\end{equation}
Unlike true many-body wavefunctions, $\psi_i(r)$ are auxiliary quantities. There are many approximations involved in solving  Kohn-Sham equations within the DFT scheme particularly modeling the exchange-correlation potential. A popular scheme is the local density approximation (LDA)\cite{PhysRev.136.B864}, which is purely dependent on the local density $\rho(r)$. Another approach is the generalized gradient approximation (GGA)\cite{PhysRevB.33.8800, PhysRevLett.77.3865} that includes both local density and its gradient.    

\begin{figure}
    \centering
    \includegraphics[width=0.8\textwidth]{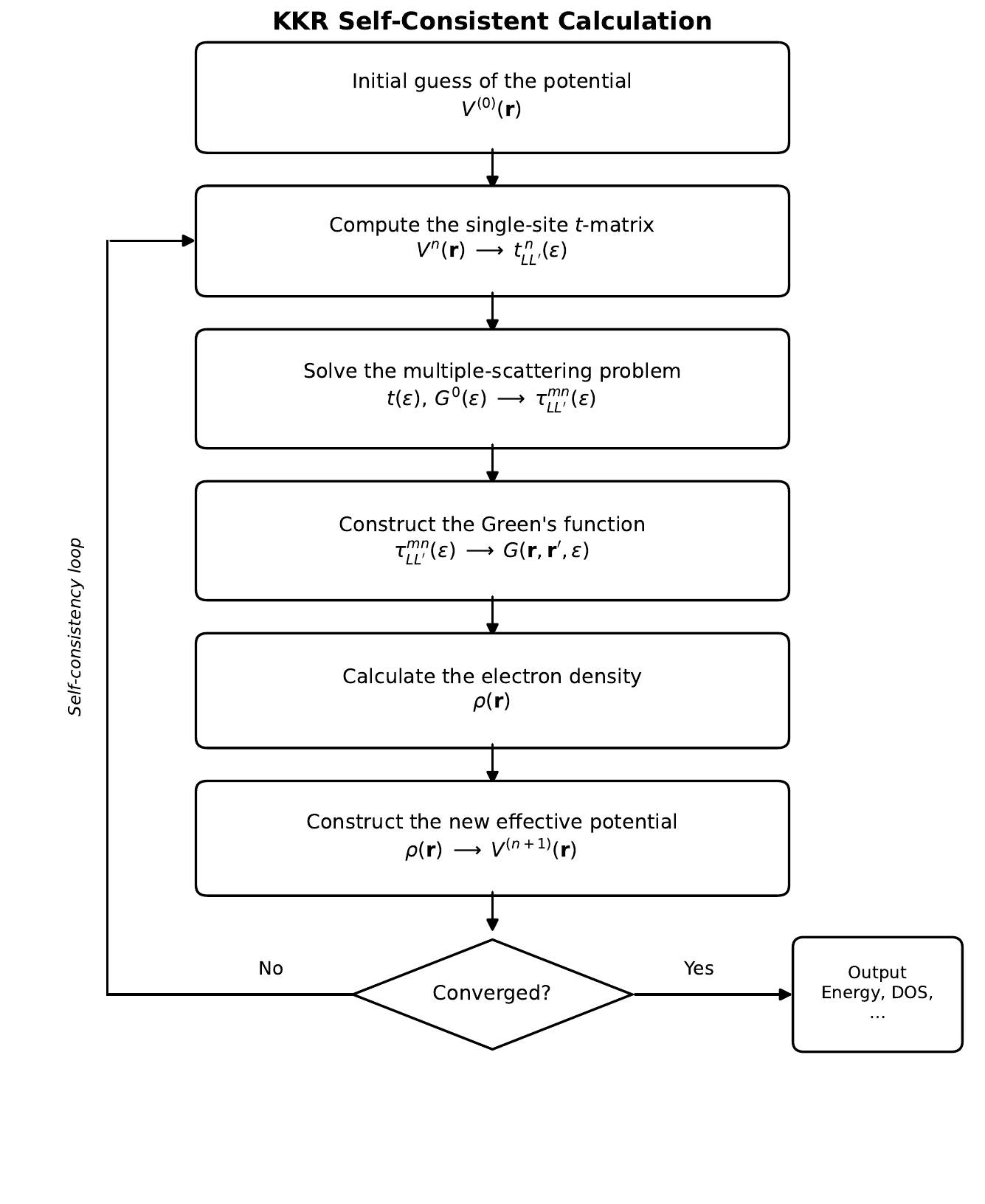}
    \caption{Flowchart of the self-consistent KKR multiple-scattering formalism.}
\label{fig:KKRselfconsistency}
\end{figure}

Popular plane-wave DFT codes like VASP, Quantum Espresso, and all-electron DFT codes like WIEN2k consider 
Eq.\ref{eq:KS} as an eigenvalue problem, whereas MST avoids the explicit diagonalization of the Hamiltonian to obtain the Kohn–Sham orbitals by adopting Green's function formalism. It is based on the scattering picture where the system is divided into non-overlapping atomic cells where each cell has an atom at its center. The total effective potential is assumed to be the sum of the cell potentials
\begin{equation}
v_n(r_n) = \begin{cases}
V_{\text{eff}} & \text{if } r \in \Omega_n \\
0 & \text{otherwise}
\end{cases}   
\end{equation}
with $r_n=r-R_n$, where $R_n$ is the atomic-site lattice vector and $\Omega_n$ is the nth cell's volume. Each of these cells is considered an electron scattering center. All these scattering events can be described by the scattering path operator. For example, the scattering path operator $\tau^{nm}(\epsilon)$ yields all possible scattering events of electron states with energy $\epsilon$ between atomic sites $n$ and $m$ and the corresponding scattering path matrix can be given as 
\begin{align}
\underline{\tau^{nm}} & = \underline{t^n} \delta_{nm} + \underline{t^n} {\underline{G}}^{nm}_0 (1-\delta_{nm}) \underline{t}^{m} + \underline{t}^{n} \sum_{k \neq n} \underline{G}^{nk}_{0} \underline{t}^{k} \underline{G}^{km}_{0} (1-\delta_{km}) \underline{t}_{m} \nonumber \\
& + \underline{t}^{n} \sum_{k \neq n} \underline{G}^{nk}_{0} \underline{t}^{k} \sum_{j \neq k} \underline {G}^{kj}_0 \underline{t}^{j} \underline{G}^{jm}_0 (1-\delta_{jm}) \underline{t}^{m} + ........ \nonumber \\
& = \underline{t}^{n} \delta_{nm} + \underline{t}^{n} \sum_{k \neq n} \underline{G}^{nk}_0 \underline {\tau}^{km}
\end{align}
where underline stands for matrices with respect to angular momentum index L, $\underline{t}^{n}(\epsilon)$ is the single site scattering t-matrix corresponding to the site n, and $\underline{G}^{nm}_0 (\epsilon)$ is the free-electron propagator matrix in the angular momentum representation, also known as KKR structure constant matrix describing the propagation of free electron with energy $\epsilon$ from site $n$ to site $m$. 

We can construct single-site Green’s function at cell n from the scattering path matrix as given below 
\begin{align}
G(\mathbf{r_n}, r_{n^\prime}; \epsilon) & = \sum_{LL^\prime} Z^{n}_{L}(\mathbf{r_n};\epsilon) \tau^{nn^\prime}_{LL^\prime}(\epsilon) Z^{n*}_{L^\prime}(\mathbf{r_n}; \epsilon) \nonumber \\
& - \sum_{L} Z^{n}_{L}(\mathbf{r_{n^\prime}}; \epsilon) J^{n*}_{L}(r_{n^\prime};\epsilon)
\label{eq:solngf}
\end{align}
where L is the composite index for angular momentum quantum number $l$ and magnetic quantum number $m$, $r_n \equiv r - R_n$ is the relative coordinate, $Z^n_L(r_n;\epsilon)$ and $J^n_L(r_n;\epsilon)$ is the regular and irregular solution of the single-site problem in cell n for momentum $L$ and energy $\epsilon$, and $\tau^{nn}_{LL^\prime}(\epsilon)$ site-diagonal matrix element of the scattering path operator $\tau^{nm}(\epsilon)$ in the angular momentum representation. The symbol $*$ signifies the complex conjugate of the associated quantities. 

The electron density at cell n from the single-site Green’s function can be computed as
\begin{equation}
\rho(r) = -\frac{1}{\pi} \text{Im} \int^{\epsilon_F}_{-\infty} G(\mathbf{r_n}, r_{n^\prime}; \epsilon) d\epsilon    
\end{equation}

An algorithm that implements the KKR for solving Eq.\ref{eq:KS} is shown in Fig\ref{fig:KKRselfconsistency}. The corresponding steps are described below: 

\begin{enumerate}
\item The computational scheme begins with an initial guess of the potential.
\item The potential is used to compute the t-matrix for each atom. 
\item  Based on the t-matrix and free electron propagator, we compute the multiple scattering path-matrix.  
\item The multiple scattering path-matrix can be used to compute a single-site Green's function. 
\item The electron density can be obtained from the single-site Green's function. 
\item The new potential corresponding to the electron density is computed. 
\item The computed new potential is compared to the old potential to satisfy the convergence requirement. If $|V_{\text{new}} - V_{\text{old}}| < \eta_{conv}$, where $\eta_{conv}$ is a small number as $10^{-5}$, then we accept the solution and the calculation stops. If this difference between old and new potential does not achieve the desired convergence, we go back to step 2, and the calculations continue until the desired convergence is reached.      
\end{enumerate}

In our calculations, we consider the diamond-cubic structure of Si, in which Ge and Sn atoms are randomly distributed over the substitutional lattice sites to model the ternary Si$_{1-(x+y)}$Ge$_y$Sn$_x$ alloy. The exchange-correlation energy is treated within the local density approximation (LDA). Brillouin-zone integrations are performed using a $12\times12\times12$ Monkhorst–Pack $\mathbf{k}$-point mesh \cite{PhysRevB.13.5188}. For the density of states (DOS) calculations, a denser $48\times48\times48$ $\mathbf{k}$-point mesh is employed to ensure improved resolution of the electronic structure. All calculations presented in this work are performed within the nonmagnetic framework, neglecting spin–orbit coupling (SOC). The single-site scattering problem is solved by expanding the scattering matrices in spherical harmonics up to an angular momentum cutoff of $l_{\mathrm{max}}=4$. Self-consistency in the electronic structure calculations is achieved using Broyden's mixing scheme applied to the effective potential, with a mixing parameter of 0.01. The convergence criteria for the total energy and effective potential are set to $5\times10^{-6}$ Ry and $1\times10^{-7}$, respectively. The equilibrium lattice parameters of all the systems investigated in this work were determined by fitting the calculated total energies as a function of volume to the third-order Birch–Murnaghan equation of state, expressed as

\[
E(V) = E_0 +
\frac{9V_0B_0}{16}
\left[
B_0'
\left\{
\left(\frac{V_0}{V}\right)^{2/3}-1
\right\}^{3}
+
\left\{
\left(\frac{V_0}{V}\right)^{2/3}-1
\right\}^{2}
\left\{
6-4\left(\frac{V_0}{V}\right)^{2/3}
\right\}
\right],
\label{eq:birch_murnaghan}
\]

where \(E_0\), \(V_0\), \(B_0\), and \(B_0'\) represent the equilibrium total energy, equilibrium volume, bulk modulus, and its first pressure derivative, respectively. The optimized lattice parameters were subsequently obtained from the corresponding equilibrium volumes \(V_0\). In addition to determining the equilibrium structural properties, the equation-of-state fitting provides the bulk modulus \(B_0\), which characterizes the resistance of a material to uniform compression. A larger bulk modulus indicates greater resistance to volumetric deformation, whereas a smaller value signifies higher compressibility. This approach enables a consistent determination of the equilibrium lattice parameters and bulk moduli across the elemental and alloy compositions considered in this study.

\begin{table*}[t]
\caption{\label{tab:benchmark}
Comparison of the equilibrium lattice parameter ($a_0$) and
bulk modulus ($B_0$) of Si, Ge, and Sn in their diamond FCC structure
obtained using KKR full-potential calculations.
Experimental measurements and previously reported theoretical
results are included where available.
A dash indicates that no corresponding value is reported to the best of our knowledge.
}
\centering
\renewcommand{\arraystretch}{1.25}
\setlength{\tabcolsep}{6pt}

\begin{tabularx}{\textwidth}{
    l
    c
    c
    c
    c
    >{\raggedright\arraybackslash}X
}
\toprule
\textbf{Property}
& \textbf{KKR-MT}
& \textbf{KKR-ASA}
& \textbf{KKR-FP}
& \textbf{Experiment}
& \textbf{Previous calculations} \\
\midrule

\multicolumn{6}{l}{\textbf{Si}} \\
\addlinespace[3pt]

$a_0$ (\AA)
& 5.95
& 5.03
& \textbf{5.433}
& 5.431~\cite{10.1063/1.1702202}
& 5.467~\cite{PhysRevLett.110.165503},
  5.402~\cite{PhysRevB.60.5202},
  5.31~\cite{10.1063/1.3618671} \\

\addlinespace[3pt]

$B_0$ (GPa)
& ---
& ---
& \textbf{94.9}
& 97.88~\cite{10.1063/1.1702809},
  99~\cite{PhysRevB.26.5668}
& 96~\cite{PhysRevB.51.9560},
  97.9~\cite{McSkimin1964} \\

\midrule

\multicolumn{6}{l}{\textbf{Ge}} \\
\addlinespace[3pt]
$a_0$ (\AA)
& ---
& ---
& \textbf{5.73}
& \textbf{5.652}\cite{PhysRevB.67.113306}
& \textbf{5.710}\cite{10.1021/acs.jpcc.5c05816}, \textbf{5.769}\cite{10.1021/acs.jpcc.5c05816}, \textbf{5.676}\cite{10.1021/acs.jpcc.5c05816}
 \\

\addlinespace[3pt]

$B_0$ (GPa)
& ---
& ---
& \textbf{70.86}
& \textbf{75}\cite{PhysRevB.67.113306}
& 72\cite{10.1021/acs.jpcc.5c05816} \\

\midrule

\multicolumn{6}{l}{\textbf{Sn}} \\
\addlinespace[3pt]
$a_0$ (\AA)
& ---
& ---
& \textbf{6.7}
& \textbf{6.483}\cite{PhysRevLett.14.554}
& \textbf{6.65}\cite{sestak2026accurate}
 \\

\addlinespace[3pt]

$B_0$ (GPa)
& ---
& ---
& \textbf{39.19}
& \textbf{53}\cite{PhysRevB.3.1243}
&  \textbf{36}\cite{PhysRevB.80.115205}, \textbf{45}\cite{PhysRevB.80.115205} \\

\bottomrule
\end{tabularx}

\end{table*}

\begin{table*}[t]
\caption{\label{tab:ternary}
Comparison of the equilibrium lattice parameter ($a_0$),
bulk modulus ($B_0$), and mixing enthalpy
($\Delta H_{\mathrm{mix}}$) of
Si$_{1-(x+y)}$Ge$_y$Sn$_x$ alloys obtained from
full-potential Korringa--Kohn--Rostoker (KKR-FP)
calculations. The lattice parameters estimated using
Vegard's law are included for comparison.
A dash indicates that the corresponding value is
not available.
}
\centering

\renewcommand{\arraystretch}{1.35}
\setlength{\tabcolsep}{12pt}

\begin{tabularx}{\textwidth}{
    >{\raggedright\arraybackslash}X
    >{\centering\arraybackslash}p{3.0cm}
    >{\centering\arraybackslash}p{3.5cm}
}
\toprule
\textbf{Property}
& \textbf{KKR-FP}
& \textbf{Vegard's law} \\
\midrule

\multicolumn{3}{l}{
\textbf{Si$_{0.5}$Ge$_{0.375}$Sn$_{0.125}$}
} \\
\addlinespace[3pt]

$a_0$ (\AA)
& \textbf{5.78}
& 5.705 \\

$B_0$ (GPa)
& \textbf{88.18}
& --- \\

$\Delta H_{\mathrm{mix}}$ (meV/atom)
& \textbf{+180.92}
& --- \\

\addlinespace[5pt]
\midrule

\multicolumn{3}{l}{
\textbf{Si$_{0.375}$Ge$_{0.25}$Sn$_{0.375}$}
} \\
\addlinespace[3pt]

$a_0$ (\AA)
& \textbf{6.11}
& 6.0 \\

$B_0$ (GPa)
& \textbf{58.23}
& --- \\

$\Delta H_{\mathrm{mix}}$ (meV/atom)
& \textbf{+346.32}
& --- \\

\addlinespace[5pt]
\midrule
\addlinespace[5pt]

\multicolumn{3}{l}{
\textbf{Si$_{0.5}$Sn$_{0.5}$}
} \\
\addlinespace[3pt]

$a_0$ (\AA)
& \textbf{6.14}
& 6.08 \\

$B_0$ (GPa)
& \textbf{58.84}
& --- \\

$\Delta H_{\mathrm{mix}}$ (meV/atom)
& \textbf{+237.87}
& --- \\

\bottomrule
\end{tabularx}

\end{table*}


\begin{figure*}[t]
    \centering

    \includegraphics[width=0.47\textwidth]{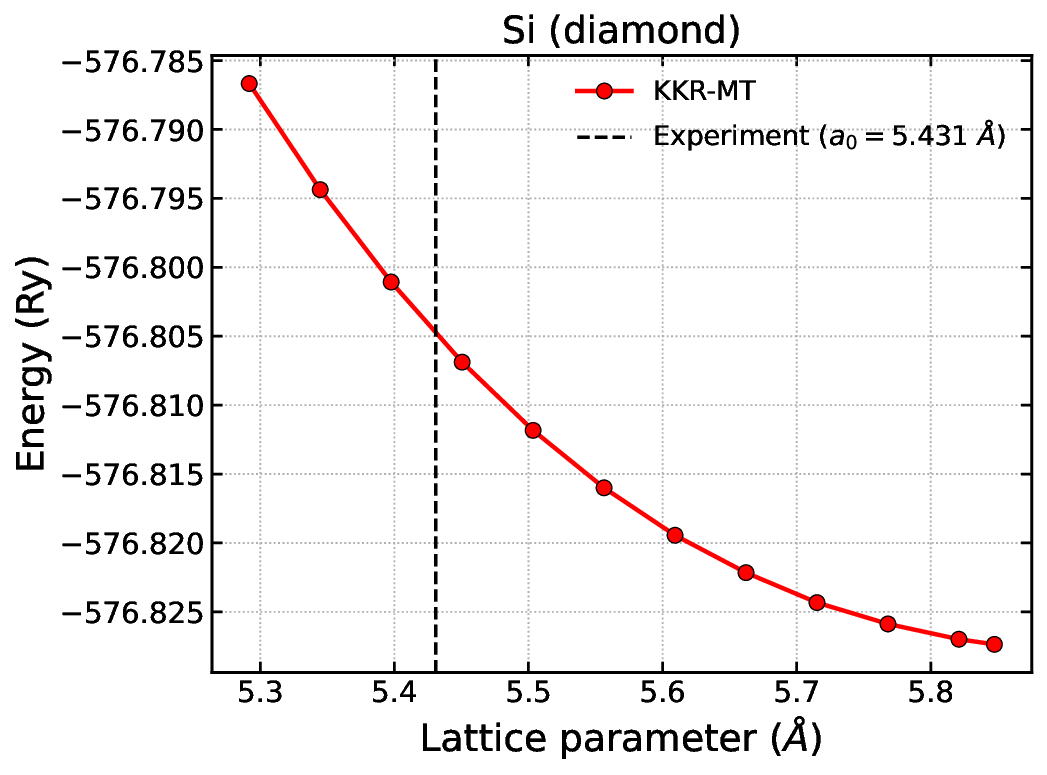}
    \hfill
    \includegraphics[width=0.47\textwidth]{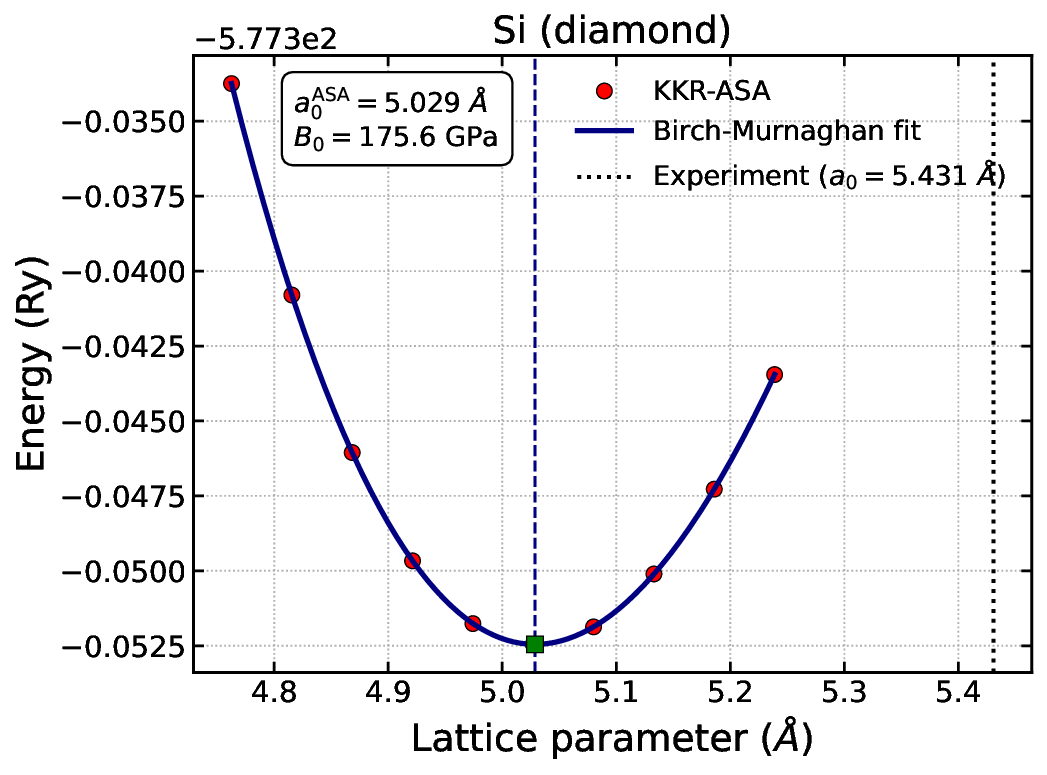}

    \par\vspace{0.3cm}

    \includegraphics[width=0.47\textwidth]{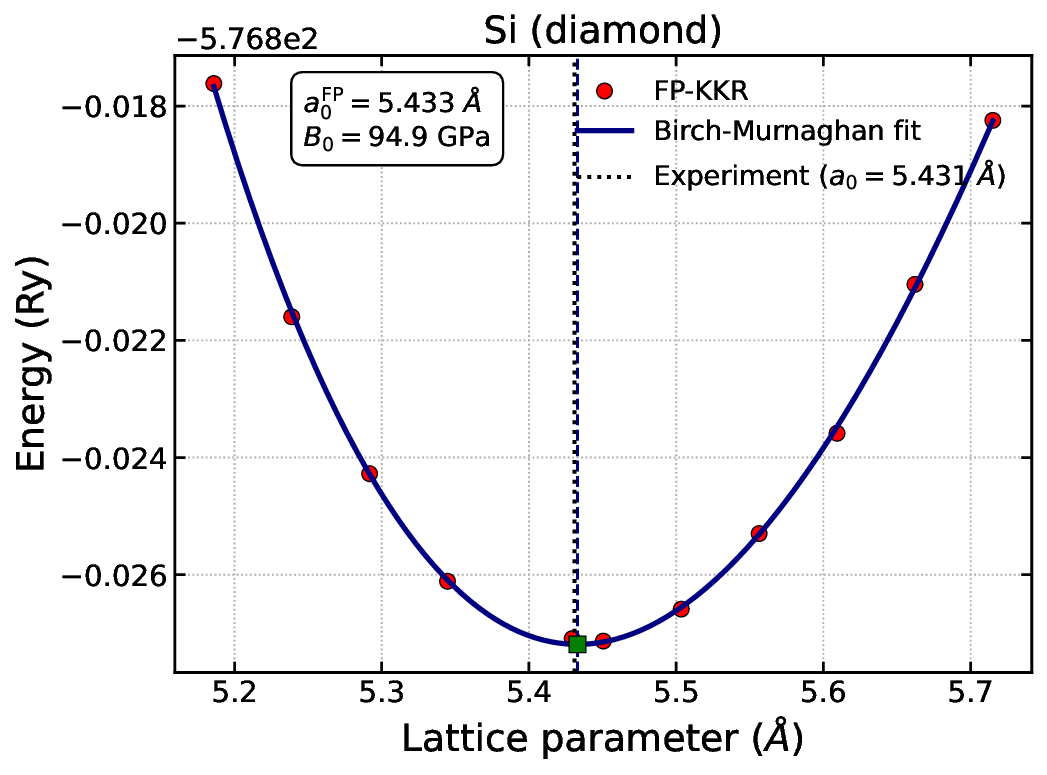}

    \caption{
    Total energy per atom as a function of lattice parameter
    for diamond-structured Si calculated using
    (a) the muffin-tin approximation (KKR-MT, upper-left panel),
    (b) the atomic sphere approximation (KKR-ASA, upper-right panel),
    and (c) the full-potential KKR method (KKR-FP, lower panel).
    The KKR-ASA and KKR-FP total-energy curves are fitted
    to the third-order Birch--Murnaghan equation of state.
    The experimental equilibrium lattice parameter is
    indicated by the vertical dashed lines.
    }
    \label{fig:Si_diffpot}
\end{figure*}

\begin{figure}
    \centering
    \includegraphics[width=0.48\textwidth]{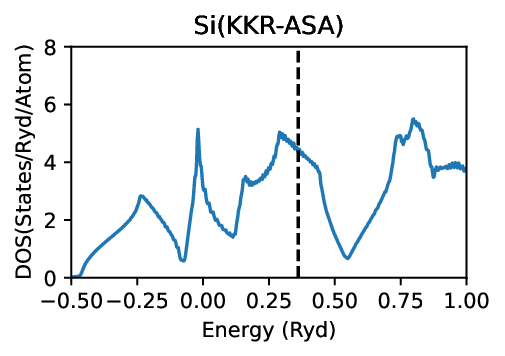}
    \includegraphics[width=0.48\textwidth]{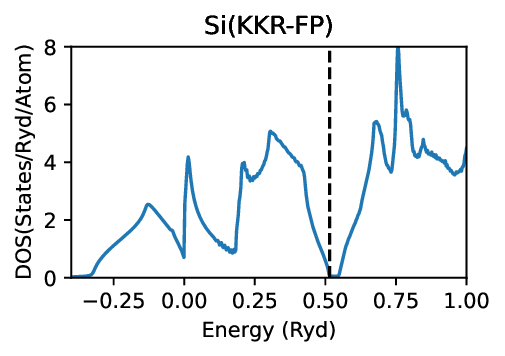}
    \caption{Density of States for Diamond-Si obtained from KKR-ASA (left panel) and KKR-FP (right panel).}
\label{fig:SiDOS}
\end{figure}


\begin{figure*}[t]
    \centering
    \includegraphics[width=0.45\textwidth]{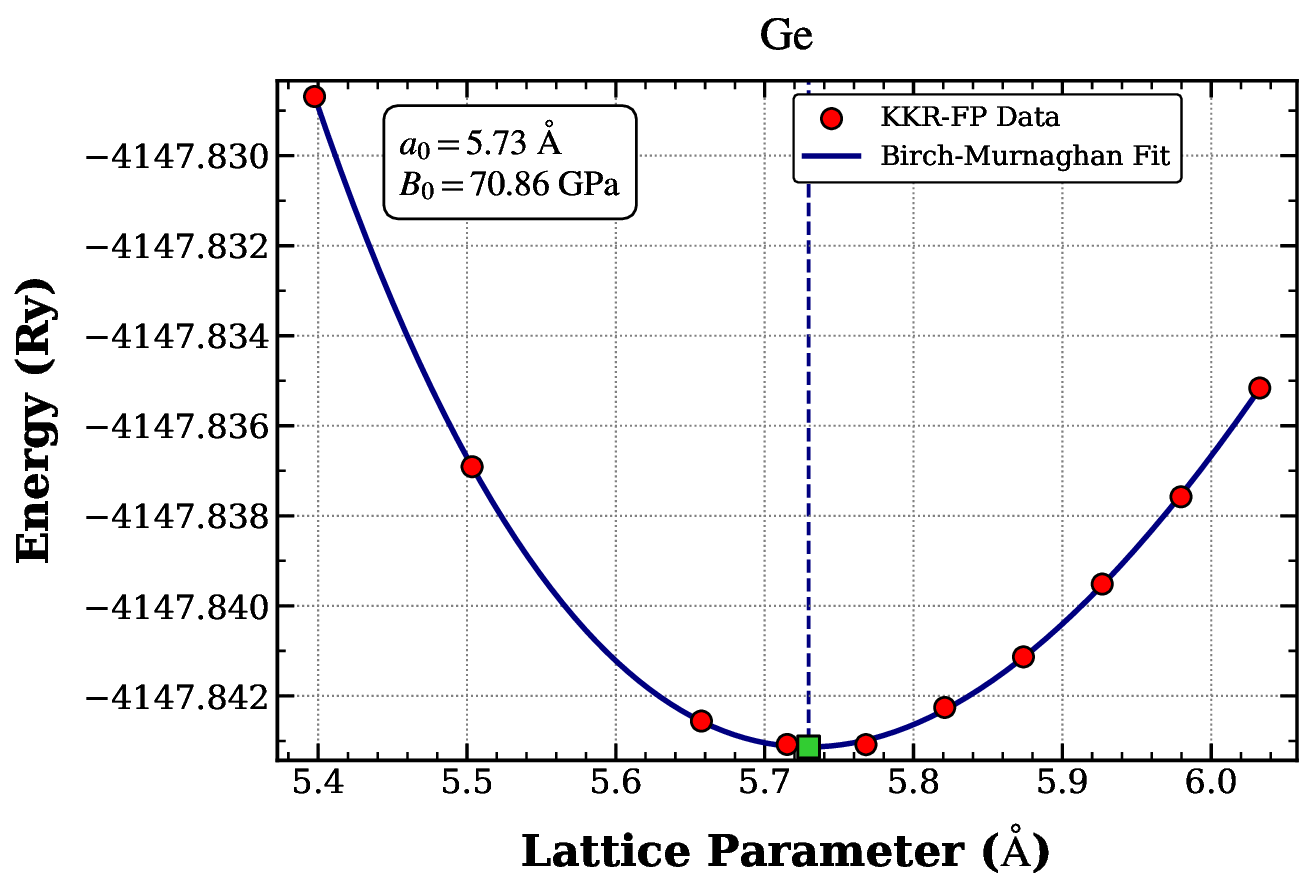}
    \includegraphics[width=0.45\textwidth]{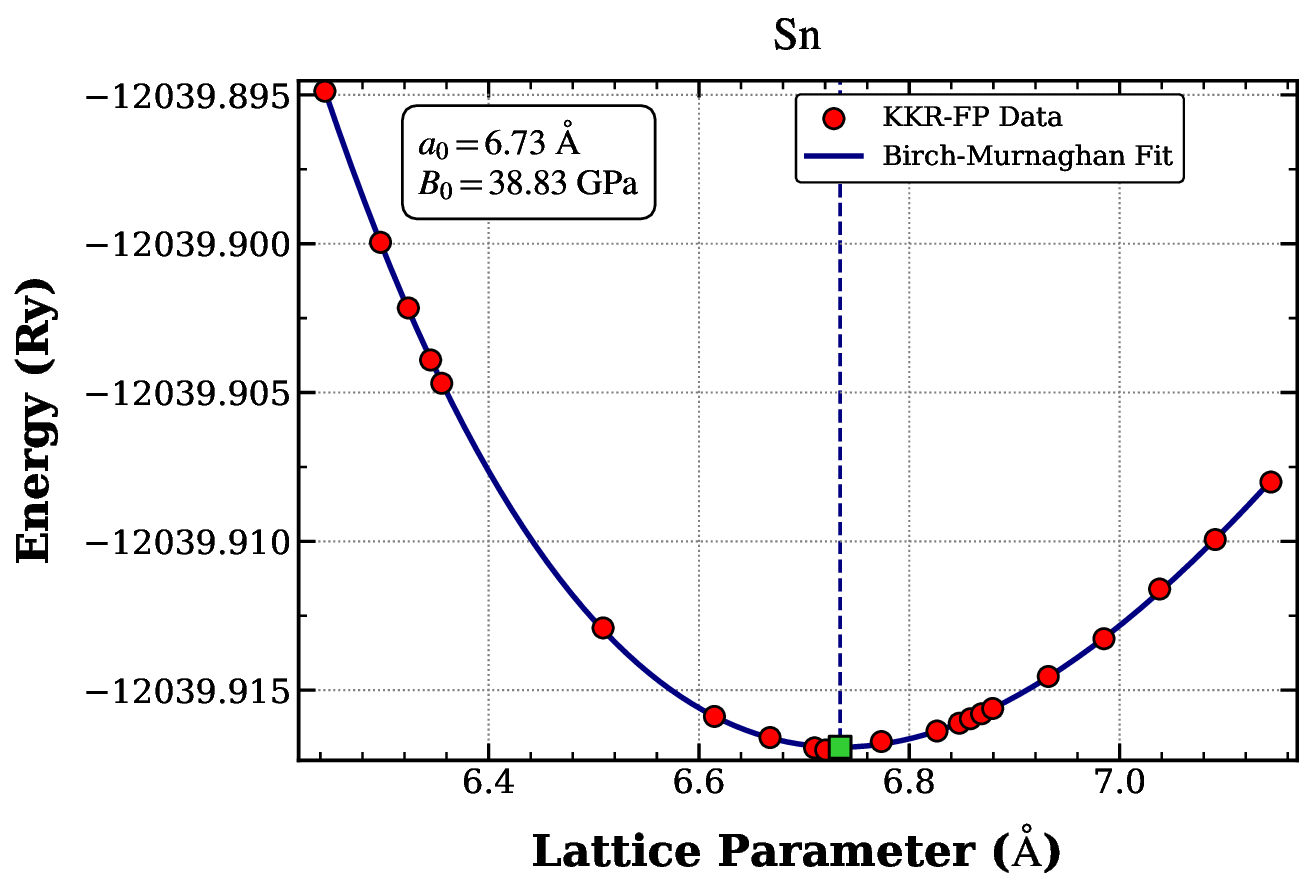}
    \caption{
    Total energy per atom as a function of lattice parameter
    calculated using the full-potential KKR method for Ge (left panel) and Sn(right panel) 
    The solid lines represent fits to the third-order
    Birch--Murnaghan equation of state.
    }
    \label{fig:EVGesepTin}
\end{figure*}

\begin{figure*}[t]
    \centering

    \includegraphics[width=0.5\textwidth]{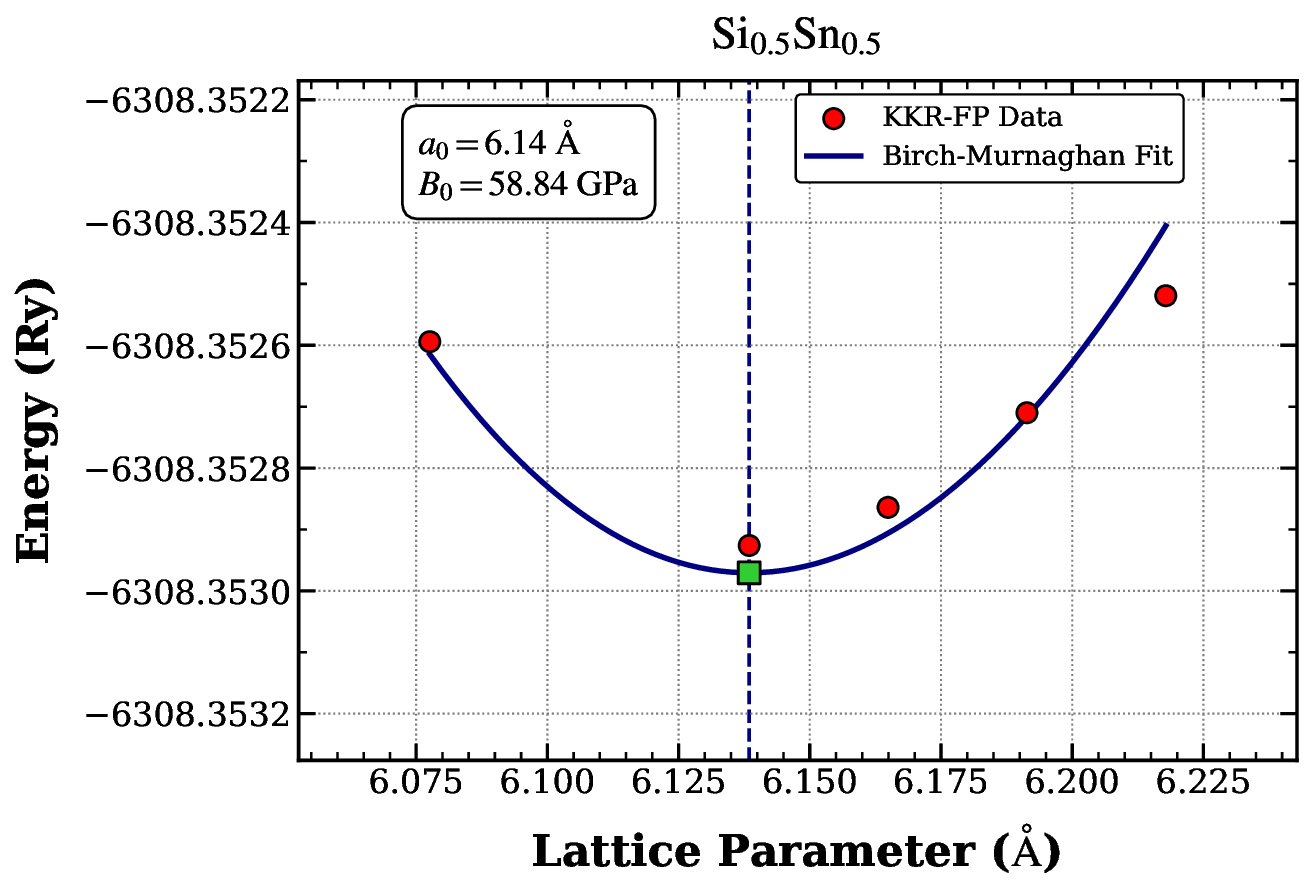}
    \hfill
    \includegraphics[width=0.5\textwidth]{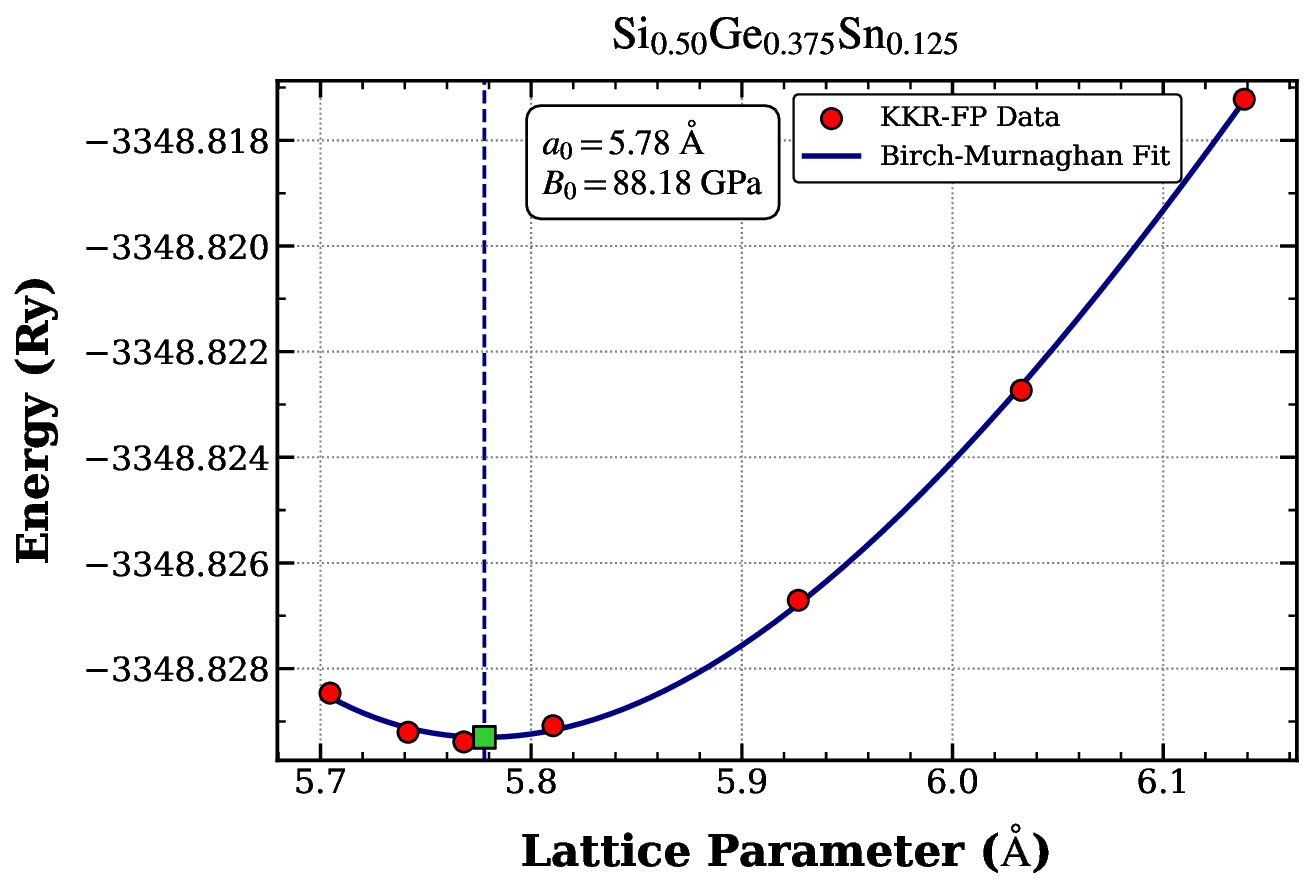}

    \par\vspace{0.3cm}

    \includegraphics[width=0.5\textwidth]{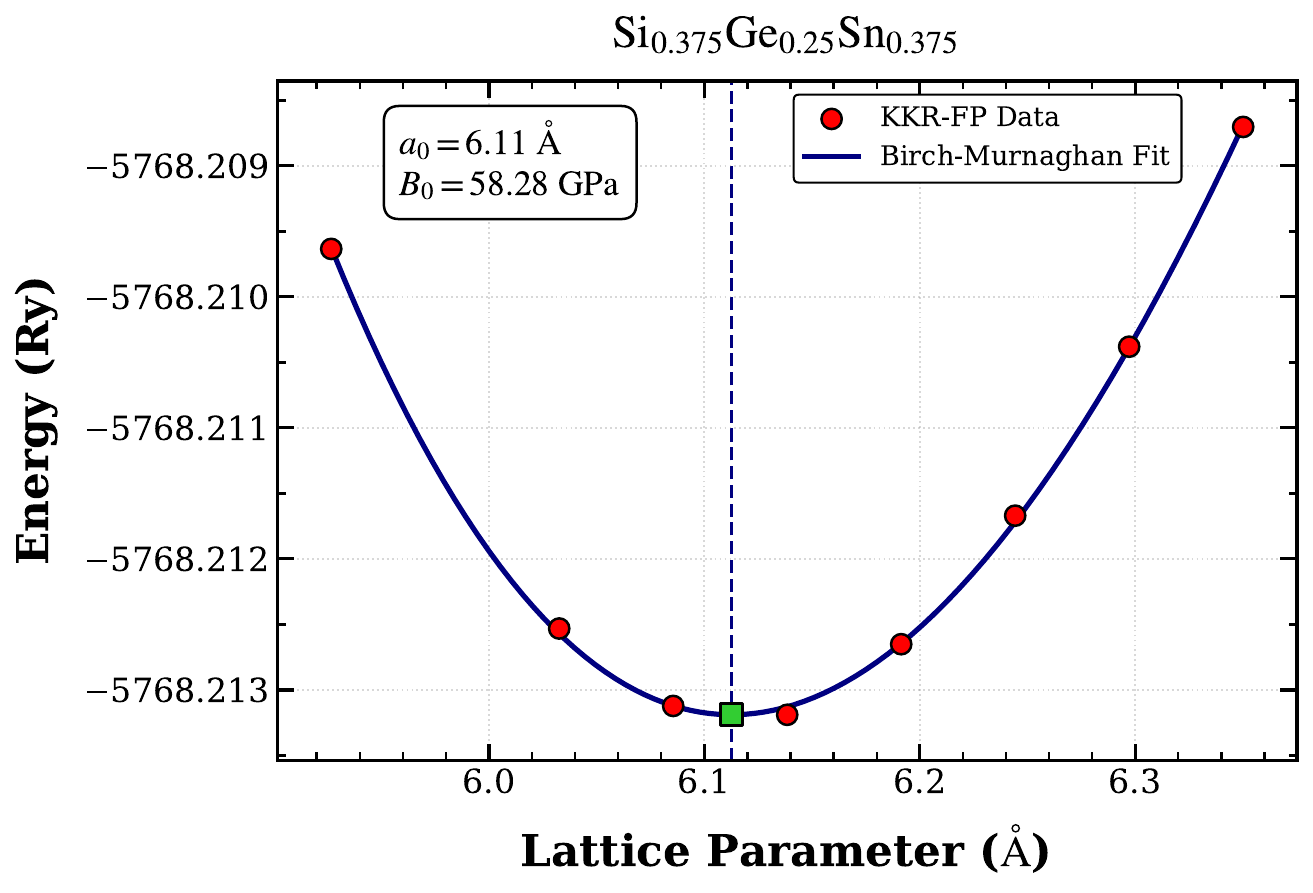}

    \caption{
    Total energy per atom as a function of lattice parameter
    calculated using the full-potential KKR method for
    (a) Si$_{0.5}$Sn$_{0.5}$,
    (b) Si$_{0.5}$Ge$_{0.375}$Sn$_{0.125}$,
    and (c) Si$_{0.375}$Ge$_{0.25}$Sn$_{0.375}$ alloys.
    The solid lines represent fits to the third-order
    Birch--Murnaghan equation of state.
    }
    \label{fig:EVSiGeSn}
\end{figure*}

\section{Results and Discussions}
\label{sec:results}

\subsection{\textbf{Benchmarking the full-potential KKR method for diamond-structured Si, Ge}}

We begin by systematically evaluating the performance of three different treatments of the effective crystal potential within the KKR framework: the muffin-tin approximation (KKR-MT), the atomic sphere approximation (KKR-ASA), and the full-potential approach (KKR-FP). Diamond-structured Si provides a particularly suitable benchmark for this comparison because of its open crystal structure and strongly directional covalent bonding, for which an accurate description of the nonspherical contributions to the effective potential is expected to be important.

Figure \ref{fig:Si_diffpot} (upper-left panel) presents the calculated total energy of diamond-structured Si as a function of the lattice parameter within the KKR-MT approximation. Interestingly, the calculated energy does not exhibit a well-defined minimum near the experimental equilibrium lattice parameter. Instead, the total energy decreases monotonically with increasing lattice parameter over the investigated range, preventing a reliable determination of the equilibrium structure. This behavior highlights the limitations of the spherical muffin-tin representation for open, covalently bonded materials such as diamond-structured Si. Our observation is consistent with the earlier analysis of Williams \cite{PhysRevB.1.3417}, who emphasized the importance of the interstitial region in determining the accuracy of muffin-tin-based calculations. Within the MTA, the effective potential is assumed to be spherically symmetric inside non-overlapping atomic spheres and constant throughout the interstitial region. As pointed out by Williams \cite{PhysRevB.1.3417}, the constant-potential approximation affects approximately 25\% of the crystal volume in closely packed structures, compared with nearly 66\% in the open diamond structure. Consequently, a substantial fraction of the crystal potential is represented by a constant in diamond-structured Si, neglecting important spatial variations associated with covalent bonding. 

To examine whether a more complete representation of the crystal volume improves the description of the equilibrium structure, we next consider the KKR-ASA approach. In contrast to the KKR-MT results, the KKR-ASA calculations yield a well-defined minimum in the total-energy curve, as shown in the upper-right panel of Fig. \ref{fig:Si_diffpot}. By fitting the calculated energies to the Birch--Murnaghan equation of state, we obtain an equilibrium lattice parameter of $a_0=5.03$ \AA, which is considerably smaller than the experimental value of $5.431$ \AA. This corresponds to an underestimation of approximately 7.4\%. Previous KKR-ASA calculations by Kotani \textit{et al}. \cite{PhysRevB.54.16502} investigated the electronic band structure of diamond-structured Si using the experimental crystal structure rather than determining its equilibrium lattice parameter through total-energy minimization. Our results demonstrate that, although the ASA provides a substantial improvement over the MTA by recovering a well-defined equilibrium structure, the resulting lattice parameter remains significantly underestimated. This discrepancy suggests that the spherical representation of the effective potential, even with space-filling atomic spheres, is insufficient to accurately describe the equilibrium structural properties of diamond-structured Si. These limitations motivate the use of a full-potential treatment in which the nonspherical contributions to the effective potential are explicitly retained.

We therefore turn to the full-potential KKR approach. The calculated total energy of diamond-structured Si as a function of the lattice parameter is presented in the lower panel of Fig. \ref{fig:Si_diffpot}. In contrast to the results obtained using the spherical-potential approximations, the KKR-FP calculations exhibit a well-defined energy minimum in close agreement with the experimentally observed equilibrium structure. Fitting the calculated energies to the Birch--Murnaghan equation of state yields an equilibrium lattice parameter of $a_0=5.433$ \AA, in excellent agreement with the experimental value of $5.431$ \AA \cite{10.1063/1.1702202}. The relative deviation from experiment is only approximately 0.04\%, demonstrating a substantial improvement over the KKR-ASA result. This agreement highlights the importance of retaining the nonspherical character of the effective potential when describing the structural properties of open, covalently bonded materials.

To further assess the accuracy of our full-potential calculations, we compare the calculated equilibrium lattice parameter with previously reported theoretical and experimental values, as summarized in Table \ref{tab:benchmark}. The equilibrium lattice parameter of diamond-structured Si has been extensively investigated using various first-principles electronic-structure methods, with the predicted values depending on the computational framework, the exchange-correlation functional, and the numerical approximations employed. Of particular relevance is the earlier full-potential KKR study by Asato \textit{et al}. \cite{PhysRevB.60.5202}, which reported an equilibrium lattice parameter of $5.402$ \AA\ within the local density approximation (LDA). Our calculated value of $5.433$ \AA, obtained using the same class of exchange-correlation functional, is reasonably consistent with this earlier result. Similarly, full-potential linear muffin-tin orbital (FP-LMTO) calculations reported an equilibrium lattice parameter of $5.41$ \AA\ \cite{PhysRevB.40.2009}. The corresponding deviations from the experimental value are approximately 0.53\% and 0.39\%, respectively, compared with only 0.04\% in our calculations. Although such differences may arise from variations in computational settings and numerical implementations, the close agreement between our KKR-FP result and the experimental equilibrium lattice parameter provides an encouraging benchmark for the full-potential implementation available in the MuST package.

In addition to the equilibrium lattice parameter, we evaluate the bulk modulus, $B_0$, which characterizes the resistance of a material to uniform compression and provides an additional measure of the accuracy of the calculated equation of state. From the Birch--Murnaghan fit to the KKR-FP total-energy data, we obtain a bulk modulus of $B_0=94.9$ GPa for diamond-structured Si. This value is in good agreement with the experimentally reported values of $97.88$ GPa \cite{10.1063/1.1702809} and $99$ GPa \cite{PhysRevB.26.5668}, as well as the previously reported theoretical value of $96$ GPa \cite{PhysRevB.51.9560}. The simultaneous agreement of the equilibrium lattice parameter and bulk modulus with available reference values provides further support for the reliability of the calculated structural energetics. Taken together, these results demonstrate that the full-potential treatment substantially improves the description of the equilibrium structural properties of diamond-structured Si compared with the spherical-potential approximations considered here.

Having established the performance of the different potential approximations in describing the equilibrium structural properties, we next examine their influence on the electronic density of states (DOS). Figure \ref{fig:SiDOS} compares the DOS of diamond-structured Si obtained using the KKR-ASA and KKR-FP approaches. As shown in the left panel, the KKR-ASA calculations yield a finite density of states at the Fermi level, suggesting metallic behavior within this approximation. This result is inconsistent with the experimentally established semiconducting character of diamond-structured Si and indicates that the spherical-potential treatment employed in the present ASA calculations does not adequately reproduce its electronic structure. In contrast, the KKR-FP DOS, shown in the right panel of Fig. \ref{fig:SiDOS}, exhibits a pronounced energy gap around the Fermi level, consistent with the semiconducting nature of Si. The recovery of a semiconducting DOS provides additional evidence that the full-potential treatment improves the description of the electronic structure.

It is important to emphasize, however, that reproducing the semiconducting character of Si does not necessarily imply an accurate prediction of its experimental band gap. The present calculations employ the LDA exchange-correlation functional, which is known to underestimate semiconductor band gaps because of limitations inherent in the conventional Kohn--Sham description. A more quantitative treatment of the band gap may therefore require improved exchange-correlation approximations or many-body corrections. Nevertheless, the structural and electronic benchmarks presented here demonstrate the advantages of the full-potential KKR framework for diamond-structured Si. These findings provide the basis for extending the present methodology to Si-based alloys, where differences in atomic size, chemical bonding, and local environments may further enhance the importance of an accurate representation of the effective crystal potential.

Having established the performance of the full-potential KKR method for diamond-structured Si, we next extend our benchmark calculations to elemental Ge, which crystallizes in the same diamond-cubic structure. This comparison provides an additional assessment of the transferability of the present computational framework to group-IV semiconductors with different atomic sizes and bonding characteristics. Similar to Si, Ge exhibits directional covalent bonding, making an accurate representation of the crystal potential important for describing its equilibrium structural properties.

The calculated total energy of diamond-structured Ge as a function of the lattice parameter, together with the corresponding Birch–Murnaghan equation-of-state fit, is presented in the left panel of Fig. \ref{fig:EVGesepTin}. As summarized in Table \ref{tab:benchmark}, our KKR-FP calculations yield an equilibrium lattice parameter of $a_0=5.73$ \AA, compared with the experimental value of $5.652$ \AA\ \cite{PhysRevB.67.113306}. The calculated lattice parameter therefore overestimates the experimental value by approximately $1.38\%$. Although the deviation is somewhat larger than that obtained for diamond-structured Si, our result remains within the range of previously reported first-principles calculations. For example, earlier theoretical studies reported equilibrium lattice parameters of $5.710$, $5.769$, and $5.676$ \AA\ \cite{10.1021/acs.jpcc.5c05816}, depending on the computational approach and exchange-correlation treatment. Our calculated value is in reasonable agreement with these theoretical predictions, providing further support for the applicability of the present KKR-FP framework to diamond-structured group-IV semiconductors.

In addition to the equilibrium lattice parameter, we calculate the bulk modulus of diamond-structured Ge from the curvature of the total-energy curve near its equilibrium volume. The resulting value, $B_0=70.86$ GPa, is in good agreement with the experimental value of approximately $75$ GPa \cite{PhysRevB.67.113306}, corresponding to an underestimation of about $5.52\%$. Our result is also consistent with the previously reported theoretical value of $72$ GPa \cite{10.1021/acs.jpcc.5c05816}. Compared with diamond-structured Si, for which we obtain $B_0=94.9$ GPa, the substantially smaller bulk modulus of Ge indicates its greater compressibility. This behavior is consistent with the larger equilibrium lattice parameter of Ge and the generally weaker resistance of its covalent bonding network to uniform compression.

As the final step of the benchmarking process, we apply KKR-FP calculations to Sn in its diamond structure, which is also known as $\alpha$-Sn. The corresponding total-energy curve and Birch–Murnaghan fit for diamond-structured $\alpha$-Sn are shown in the right panel of Fig. \ref{fig:EVGesepTin}. 
Our KKR-FP calculations find that the optimized lattice parameter for Sn is 6.73 \AA\ which is slightly larger than the experimental value (6.483 \AA)\cite{PhysRevLett.14.554}. However, previous generalized gradient approximation (GGA) based calculations show an overestimation trend compared to the experimental value as 6.65 \AA\ \cite{sestak2026accurate}. Previous LDA-based calculations underestimate the lattice parameter as 6.38 \AA\   \cite{PhysRevB.57.10421}. Calculations based on non-local functionals like SCAN and HSE yield underestimation of the lattice parameter compared to the experimental value as 6.5412  \AA\ \cite{PhysRevMaterials.5.083608} and 6.556 \AA\ \cite{znkf-3g37}. Recent LDA-based calculations, including both atomic position and lattice parameter variation, fully relaxed calculations yield 6.4785 \AA\ which is very close to the experimental value (6.483). We note that, in our optimized calculations, we only varied the lattice parameter, keeping atomic positions fixed. Overall, our KKR-FP calculations predicted a lattice parameter of 6.7 \AA\ which is in reasonable agreement with both experimental and previous calculated values.            

Taken together, the results for Si and Ge demonstrate that the full-potential KKR approach provides a consistent description of the equilibrium structural properties of these two diamond-structured group-IV semiconductors. Although the calculated lattice parameter of Ge exhibits a larger deviation from experiment than that of Si, both the equilibrium lattice parameter and bulk modulus remain in reasonable agreement with available experimental measurements and previous theoretical predictions. These benchmarks establish a useful reference for the subsequent investigation of SiGeSn alloys, where differences in atomic size and bonding characteristics among Si, Ge, and Sn are expected to influence the equilibrium lattice parameter, elastic response, and composition-dependent structural behavior.

\subsection{\textbf{Structural properties and mixing energetics of SiGeSn alloys}}

Having established the performance of the full-potential KKR method for elemental Si and Ge, we now extend our investigation to Si$_{1-(x+y)}$Ge$_y$Sn$_x$ alloys to examine the influence of chemical composition on their equilibrium structural properties and mixing energetics. In particular, we investigate the equilibrium lattice parameter ($a_0$), bulk modulus ($B_0$), and mixing enthalpy ($\Delta H_{\mathrm{mix}}$) for selected alloy compositions. The calculated results are summarized in Table \ref{tab:ternary}, together with the corresponding lattice parameters estimated using Vegard's law. These quantities provide complementary insights into the effects of chemical substitution on lattice expansion, elastic response, and the energetic favorability of alloy formation. Figure \ref{fig:EVSiGeSn} presents the calculated total energies as functions of lattice parameter for Si$_{0.5}$Sn$_{0.5}$, Si$_{0.5}$Ge$_{0.375}$Sn$_{0.125}$, and Si$_{0.375}$Ge$_{0.25}$Sn$_{0.375}$, respectively. The solid lines represent fits to the third-order Birch–Murnaghan equation of state, from which the equilibrium lattice parameters and bulk moduli are extracted.

We first examine the variation of the equilibrium lattice parameter with alloy composition. As shown in Table \ref{tab:ternary}, the calculated lattice parameter increases from $5.78$ \AA\ for Si$_{0.5}$Ge$_{0.375}$Sn$_{0.125}$ to $6.11$ \AA\ for Si$_{0.375}$Ge$_{0.25}$Sn$_{0.375}$. For comparison, the binary Si$_{0.5}$Sn$_{0.5}$ alloy exhibits an equilibrium lattice parameter of $6.14$ \AA. The overall lattice expansion is consistent with the incorporation of larger Sn atoms into the Si-based diamond lattice. Since Sn has a larger equilibrium atomic volume than Si and Ge, its incorporation modifies the local bonding environments and increases the equilibrium lattice parameter. However, because the relative concentrations of Si and Ge also change between the investigated ternary compositions, the observed lattice expansion reflects the combined influence of all three constituent elements rather than the Sn concentration alone.

To quantify the deviation from a linear compositional dependence, we compare the calculated equilibrium lattice parameters with those predicted by Vegard's law. For a ternary Si$_{1-(x+y)}$Ge$_y$Sn$_x$ alloy, Vegard's law is expressed as

\[
a_{\mathrm{Vegard}}(x,y)
=(1-x-y)a_{\mathrm{Si}}
+y a_{\mathrm{Ge}}
+x a_{\mathrm{Sn}},
\]

where $a_{\mathrm{Si}}$, $a_{\mathrm{Ge}}$, and $a_{\mathrm{Sn}}$ are the reference equilibrium lattice parameters of the constituent elements. As summarized in Table \ref{tab:ternary}, the KKR-FP calculations consistently predict larger equilibrium lattice parameters than the corresponding Vegard-law estimates. For Si$_{0.5}$Ge$_{0.375}$Sn$_{0.125}$, the calculated lattice parameter of $5.78$ \AA\ exceeds the Vegard-law prediction of $5.705$ \AA\ by $0.075$ \AA, corresponding to a positive deviation of approximately $1.31\%$. A larger deviation is observed for Si$_{0.375}$Ge$_{0.25}$Sn$_{0.375}$, for which the calculated lattice parameter of $6.11$ \AA\ exceeds the Vegard-law estimate of $6.00$ \AA\ by $0.11$ \AA, or approximately $1.83\%$. Similarly, the binary Si$_{0.5}$Sn$_{0.5}$ alloy exhibits a positive deviation of $0.06$ \AA\ from the Vegard-law prediction of $6.08$ \AA, corresponding to approximately $0.99\%$.

The consistently positive deviations from Vegard's law indicate that the equilibrium lattice parameters of the investigated alloys cannot be fully described by a simple linear interpolation between those of the constituent elements. Such deviations may arise from differences in atomic size, chemical bonding, and local structural environments. In particular, the incorporation of Sn into the diamond lattice introduces a substantial size mismatch with Si and Ge, which can modify the equilibrium bond lengths and generate local strain. The resulting structural response may favor an equilibrium lattice parameter larger than that predicted by linear interpolation. Interestingly, the positive deviation is more pronounced for Si$_{0.375}$Ge$_{0.25}$Sn$_{0.375}$ than for the binary Si$_{0.5}$Sn$_{0.5}$ alloy, despite the latter containing a higher Sn concentration. This observation suggests that the deviation from Vegard's law depends not only on the Sn concentration but also on the relative proportions of Si and Ge and their associated bonding environments. A more extensive investigation of the ternary composition space would be required to establish the detailed dependence of lattice-parameter bowing on alloy composition.

We next investigate the composition dependence of the bulk modulus, which characterizes the resistance of a material to uniform compression. The calculated bulk modulus decreases substantially from $88.18$ GPa for Si$_{0.5}$Ge$_{0.375}$Sn$_{0.125}$ to $58.23$ GPa for Si$_{0.375}$Ge$_{0.25}$Sn$_{0.375}$. The binary Si$_{0.5}$Sn$_{0.5}$ alloy exhibits a comparable bulk modulus of $58.84$ GPa. Relative to the lower-Sn ternary composition, the bulk modulus of Si$_{0.375}$Ge$_{0.25}$Sn$_{0.375}$ is reduced by approximately $34\%$, indicating a pronounced increase in compressibility. This reduction is consistent with the accompanying lattice expansion and reflects the changes in the bonding characteristics and equilibrium energy landscape upon alloying.

The observed reduction in bulk modulus can be qualitatively understood in terms of the structural and chemical differences among Si, Ge, and Sn. The incorporation of larger Sn atoms modifies the equilibrium bond lengths and local bonding environments within the diamond lattice, which can reduce the resistance to hydrostatic compression. In addition, the bulk modulus is directly related to the curvature of the total-energy curve near the equilibrium volume, and its reduction indicates a comparatively smaller energetic penalty for uniform volume deformation. However, the elastic response is governed by the combined effects of atomic size, bonding strength, and electronic structure, and therefore cannot be attributed solely to lattice expansion. The calculated results nevertheless demonstrate that chemical substitution substantially modifies the mechanical response of the investigated SiGeSn alloys.

Beyond the equilibrium structural properties, we evaluate the mixing enthalpy to investigate the energetic favorability of alloy formation. For the ternary Si$_{1-(x+y)}$Ge$_y$Sn$_x$ alloy, the mixing enthalpy is defined as

\[
\Delta H_{\mathrm{mix}}(x,y)
=
E_{\mathrm{Si}_{1-(x+y)}\mathrm{Ge}_y\mathrm{Sn}_x}
-(1-x-y)E_{\mathrm{Si}}
-yE_{\mathrm{Ge}}
-xE_{\mathrm{Sn}},
\]

where $E_{\mathrm{Si}_{1-(x+y)}\mathrm{Ge}_y\mathrm{Sn}_x}$ denotes the total energy per atom of the alloy, while $E_{\mathrm{Si}}$, $E_{\mathrm{Ge}}$, and $E_{\mathrm{Sn}}$ represent the corresponding elemental reference energies per atom. All energies are evaluated within a consistent computational framework.

As shown in Table \ref{tab:ternary}, the calculated mixing enthalpies are positive for all three compositions considered here. Specifically, we obtain $\Delta H_{\mathrm{mix}}=+180.92$ meV/atom for Si$_{0.5}$Ge$_{0.375}$Sn$_{0.125}$ and $+346.32$ meV/atom for Si$_{0.375}$Ge$_{0.25}$Sn$_{0.375}$. For comparison, the binary Si$_{0.5}$Sn$_{0.5}$ alloy exhibits a mixing enthalpy of $+237.87$ meV/atom. The positive values indicate that the formation of these homogeneous alloys is energetically unfavorable relative to the concentration-weighted combination of the chosen elemental reference states at zero temperature. The substantially larger mixing enthalpy obtained for Si$_{0.375}$Ge$_{0.25}$Sn$_{0.375}$ suggests a stronger energetic penalty for alloy formation at this composition. Notably, its mixing enthalpy also exceeds that of the binary Si$_{0.5}$Sn$_{0.5}$ alloy, despite containing a lower Sn concentration. This comparison indicates that the energetic cost of alloy formation depends on the combined chemical composition and bonding environments rather than solely on the Sn content.

The positive mixing enthalpies suggest an energetic driving force toward chemical segregation or phase separation relative to the selected elemental reference states. Such behavior is consistent with the challenges associated with incorporating substantial concentrations of Sn into group-IV semiconductor alloys. However, the mixing enthalpy alone does not determine the preferred decomposition pathway or establish the thermodynamic stability of an alloy against all competing phases. A complete assessment requires the construction of the formation-energy convex hull, including relevant competing phases, and the evaluation of finite-temperature free-energy contributions.

At finite temperatures, configurational entropy can favor the formation of chemically mixed phases and partially compensate for the positive enthalpy of mixing. The thermodynamic behavior of SiGeSn alloys is therefore governed by the competition between the energetic cost of mixing and the entropic stabilization associated with chemical disorder. Vibrational contributions, epitaxial strain, and nonequilibrium growth conditions may further influence the stability and experimentally accessible composition range of these alloys. Consequently, although the calculated positive mixing enthalpies indicate an energetic preference against homogeneous mixing relative to the chosen reference states, quantitative predictions of miscibility gaps, segregation temperatures, or phase boundaries require additional thermodynamic analysis.

Overall, the present full-potential KKR calculations reveal significant composition-dependent changes in the structural and energetic properties of SiGeSn alloys. The investigated compositions exhibit equilibrium lattice parameters larger than those predicted by Vegard's law, a pronounced reduction in bulk modulus relative to the lower-Sn alloy, and positive mixing enthalpies. These findings highlight the important roles of atomic-size mismatch, chemical bonding, and local structural environments in determining the properties of SiGeSn alloys. The results provide a first-principles foundation for understanding their structural behavior and motivate further investigations of alloy stability and phase separation across the ternary composition space.

\section{\label{sec:level1}Conclusions}

In summary, we have systematically investigated the structural and mechanical properties of diamond-structured group-IV semiconductors and their alloys using the recently developed full-potential Korringa--Kohn--Rostoker (KKR-FP) method implemented in the open-source MuST package. By comparing the full-potential approach with the conventional muffin-tin (MTA) and atomic sphere (ASA) approximations, we have demonstrated the importance of accurately treating the nonspherical character of the effective crystal potential, particularly in open, covalently bonded structures. For diamond-structured Si, the KKR-MTA calculations fail to produce a physically meaningful equilibrium lattice parameter over the investigated range, while the KKR-ASA approach significantly underestimates the experimental lattice parameter. In contrast, the KKR-FP calculations yield an equilibrium lattice parameter of $5.433$ \AA\ and a bulk modulus of $94.9$ GPa, in excellent agreement with the corresponding experimental values. Furthermore, the full-potential treatment successfully reproduces the semiconducting character of Si, in contrast to the metallic behavior obtained within the present KKR-ASA calculations. These results establish the improved performance of the full-potential approach for describing the structural and electronic properties of diamond-structured Si.

We have further extended the KKR-FP calculations to elemental Ge and diamond-structured $\alpha$-Sn, obtaining equilibrium structural properties in reasonable agreement with available experimental measurements and previous theoretical predictions. Building upon these benchmarks, we have investigated the composition-dependent structural properties and mixing energetics of Si$_{1-(x+y)}$Ge$_y$Sn$_x$ ternary alloys. Our calculations reveal a systematic lattice expansion across the selected compositions, accompanied by substantial changes in the bulk modulus. Notably, the calculated equilibrium lattice parameters consistently exceed the corresponding predictions of Vegard's law, indicating a positive deviation from linear compositional interpolation. These deviations highlight the influence of atomic-size mismatch, chemical bonding, and local structural environments on the equilibrium properties of SiGeSn alloys. In addition, the calculated bulk modulus decreases from $88.18$ GPa for Si$_{0.5}$Ge$_{0.375}$Sn$_{0.125}$ to $58.23$ GPa for Si$_{0.375}$Ge$_{0.25}$Sn$_{0.375}$, demonstrating a pronounced modification of the elastic response with alloy composition.

An important finding of the present study is the positive mixing enthalpy obtained for the investigated alloy compositions. In particular, the calculated mixing enthalpies of $+180.92$ and $+346.32$ meV/atom for Si$_{0.5}$Ge$_{0.375}$Sn$_{0.125}$ and Si$_{0.375}$Ge$_{0.25}$Sn$_{0.375}$, respectively, indicate an energetic penalty for homogeneous alloy formation relative to the chosen elemental reference states. A positive mixing enthalpy of $+237.87$ meV/atom is also obtained for the binary Si$_{0.5}$Sn$_{0.5}$ alloy. These results suggest an energetic tendency toward chemical segregation or phase separation, emphasizing the importance of alloy composition in determining structural and thermodynamic behavior. Nevertheless, a quantitative description of phase stability requires consideration of competing phases and finite-temperature contributions to the free energy.

Overall, the present work demonstrates the applicability of the full-potential KKR implementation in MuST to open-structured semiconductors and compositionally complex group-IV alloys. The calculated structural properties and mixing energetics provide a first-principles foundation for understanding the effects of chemical substitution in SiGeSn alloys and offer a useful perspective for interpreting experimental observations of alloy stability. Future developments will focus on extending the present framework to incorporate improved exchange-correlation treatments for more accurate electronic band-gap predictions, together with systematic investigations of composition-dependent free energies and finite-temperature phase stability. These advances will further strengthen the predictive capabilities of the KKR Green's-function approach for investigating chemically disordered semiconductor alloys.

\ack{
This work is based on open-source ab initio software package MuST \cite{Must}, a
project supported in part by NSF Office of Advanced Cyberinfrastructure and
the Division of Materials Research within the NSF Directorate of Mathematical and Physical Sciences: HT and WM acknowledge the U.S. Department of Energy, Office of Science, under award number DE-SC0025748 grant, DOE RENEW program; KMT acknowledges NSF OAC-1931445; and YW acknowledges NSF OAC-1931525. Computations in this work were performed on the Bridges-2 system at the Pittsburgh Supercomputing Center (PSC) through allocation PHY260495 from the Advanced Cyberinfrastructure Coordination Ecosystem: Services and Support (ACCESS) program, which is supported by the National Science Foundation under grants Nos. 2138259, 2138286, 2138307, 2137603, and 2138296. This research used resources of the Oak Ridge Leadership Computing Facility, which is a DOE Office of Science User Facility supported under Contract DE-AC05-00OR22725. During the preparation of this work, the authors used OpenAI (GPT-5.6) in order to search the literature, verify assumptions, correct grammar, and revise language. After using this tool/service, the authors reviewed and edited the content as needed and take full responsibility for the content of the published article.
}

\bibliographystyle{iopart-num} 
\bibliography{aipsamp}


\end{document}